\documentclass[twocolumn,preprintnumbers,amsmath,amssymb,superscriptaddress]{revtex4}

\usepackage{graphicx}
\usepackage{dcolumn}
\usepackage{bm}

\makeatletter
\def\@seccntformat#1{}
\makeatother
\renewcommand{\numberline}[1]{}

\begin{document}

\hspace{1cm} Detailed version of [Hsieh et.al., NATURE \textbf{452}, 970-974 (2008), \textbf{Submitted in November 2007}]

\title{A topological Dirac insulator in a quantum spin Hall phase (first experimental realization of a 3D Topological Insulator)}


\author{D. Hsieh}
\affiliation{Joseph Henry Laboratories of Physics, Department of Physics, Princeton
University, Princeton, NJ 08544, USA}

\author{D. Qian}
\affiliation{Joseph Henry Laboratories of Physics, Department of Physics, Princeton
University, Princeton, NJ 08544, USA}

\author{L. Wray}
\affiliation{Joseph Henry Laboratories of Physics, Department of Physics, Princeton
University, Princeton, NJ 08544, USA}

\author{Y. Xia}
\affiliation{Joseph Henry Laboratories of Physics, Department of Physics, Princeton
University, Princeton, NJ 08544, USA}
\affiliation{Princeton Center
for Complex Materials, Princeton University, Princeton NJ 08544,
USA}

\author{Y. S. Hor}
\affiliation{Department of Chemistry, Princeton University,
Princeton, NJ 08544, USA}

\author{R. J. Cava}
\affiliation{Department of Chemistry, Princeton University,
Princeton, NJ 08544, USA}

\author{M. Z. Hasan}
\affiliation{Joseph Henry Laboratories of Physics, Department of Physics, Princeton
University, Princeton, NJ 08544, USA}
\affiliation{Princeton Center
for Complex Materials, Princeton University, Princeton NJ 08544,
USA}
\affiliation{Princeton Institute for the Science and Technology of Materials (PRISM), Princeton University, Princeton NJ 08544,
USA}
\email{mzhasan@Princeton.edu}




\maketitle

\textbf{When electrons are subject to a large external magnetic
field, the conventional charge quantum Hall effect
\cite{Klitzing,Tsui} dictates that an electronic excitation gap is
generated in the sample bulk, but metallic conduction is permitted
at the boundary. Recent theoretical models suggest that certain bulk
insulators with large spin-orbit interactions may also naturally
support conducting topological boundary states in the extreme
quantum limit \cite{Kane(Graphene),Bernevig(QSHE),Sheng(QSHE)},
which opens up the possibility for studying unusual quantum
Hall-like phenomena in zero external magnetic fields
\cite{Haldane(P-anomaly)}. Bulk Bi$_{1-x}$Sb$_x$ single crystals are
predicted to be prime candidates \cite{Fu:STI1,Murukami} for one
such unusual Hall phase of matter known as the topological insulator
\cite{Fu:STI2, Moore:STI1, Roy}. The hallmark of a topological
insulator is the existence of metallic surface states that are
higher dimensional analogues of the edge states that characterize a
quantum spin Hall insulator
\cite{Kane(Graphene),Bernevig(QSHE),Sheng(QSHE),Haldane(P-anomaly),Fu:STI1,
Murukami, Fu:STI2, Moore:STI1, Roy, Bernevig:HgTe, Konig}. In
addition to its interesting boundary states, the bulk of
Bi$_{1-x}$Sb$_x$ is predicted to exhibit three-dimensional Dirac
particles \cite{Wolff, Fukuyama, Buot, Lenoir}, another topic of
heightened current interest following the new findings of
two-dimensional graphene \cite{Zhang, Novoselov, Zhou} and charge
quantum Hall fractionalization observed in pure bismuth
\cite{Behnia}. However, despite numerous transport and magnetic
measurements on the Bi$_{1-x}$Sb$_x$ family since the 1960s
\cite{Lenoir}, no direct evidence of either topological quantum
Hall-like states or bulk Dirac particles has ever been found. Here,
using incident-photon-energy-modulated angle-resolved photoemission
spectroscopy (IPEM-ARPES), we report the direct observation of
massive Dirac particles in the bulk of Bi$_{0.9}$Sb$_{0.1}$, locate
the Kramers' points at the sample's boundary and provide a
comprehensive mapping of the topological Dirac insulator's gapless
surface modes. These findings taken together suggest that the
observed surface state on the boundary of the bulk insulator is a
realization of the much sought exotic ``topological metal''
\cite{Fu:STI2, Moore:STI1, Roy}. They also suggest that this
material has potential application in developing next-generation
quantum computing devices that may incorporate ``light-like" bulk
carriers and topologically protected spin-textured edge-surface
currents.}

Bismuth is a semimetal with strong spin-orbit interactions. Its band
structure is believed to feature an indirect negative gap between
the valence band maximum at the T point of the bulk Brillouin zone
(BZ) and the conduction band minima at three equivalent L points
\cite{Lenoir,Liu} (here we generally refer to these as a single
point, L). The valence and conduction bands at L are derived from
antisymmetric (L$_a$) and symmetric (L$_s$) $p$-type orbitals,
respectively, and the effective low-energy Hamiltonian at this point
is described by the (3+1)-dimensional relativistic Dirac equation
\cite{Wolff, Fukuyama, Buot}. The resulting dispersion relation,
$E(\vec{k}) = \pm \sqrt{ {(\vec{v} \cdot \vec{k})}^2 + \Delta^2}
\approx \vec{v} \cdot \vec{k}$, is highly linear owing to the
combination of an unusually large band velocity $\vec{v}$ and a
small gap $\Delta$ (such that $\lvert \Delta / \lvert \vec{v} \rvert
\rvert \approx 5 \times 10^{-3} $\AA$^{-1}$) and has been used to
explain various peculiar properties of bismuth \cite{Wolff,
Fukuyama, Buot}. Substituting bismuth with antimony is believed to
change the critical energies of the band structure as follows (see
Fig.1). At an Sb concentration of $x \approx 4\%$, the gap $\Delta$
between L$_a$ and L$_s$ closes and a massless three-dimensional (3D)
Dirac point is realized. As $x$ is further increased this gap
re-opens with inverted symmetry ordering, which leads to a change in
sign of $\Delta$ at each of the three equivalent L points in the BZ.
For concentrations greater than $x \approx 7\%$ there is no overlap
between the valence band at T and the conduction band at L, and the
material becomes an inverted-band insulator. Once the band at T
drops below the valence band at L, at $x \approx 8\%$, the system
evolves into a direct-gap insulator whose low energy physics is
dominated by the spin-orbit coupled Dirac particles at L
\cite{Fu:STI1, Lenoir}.

\begin{figure*}
\includegraphics[scale=0.75,clip=true, viewport=0.0in 0in 8.5in 7.4in]{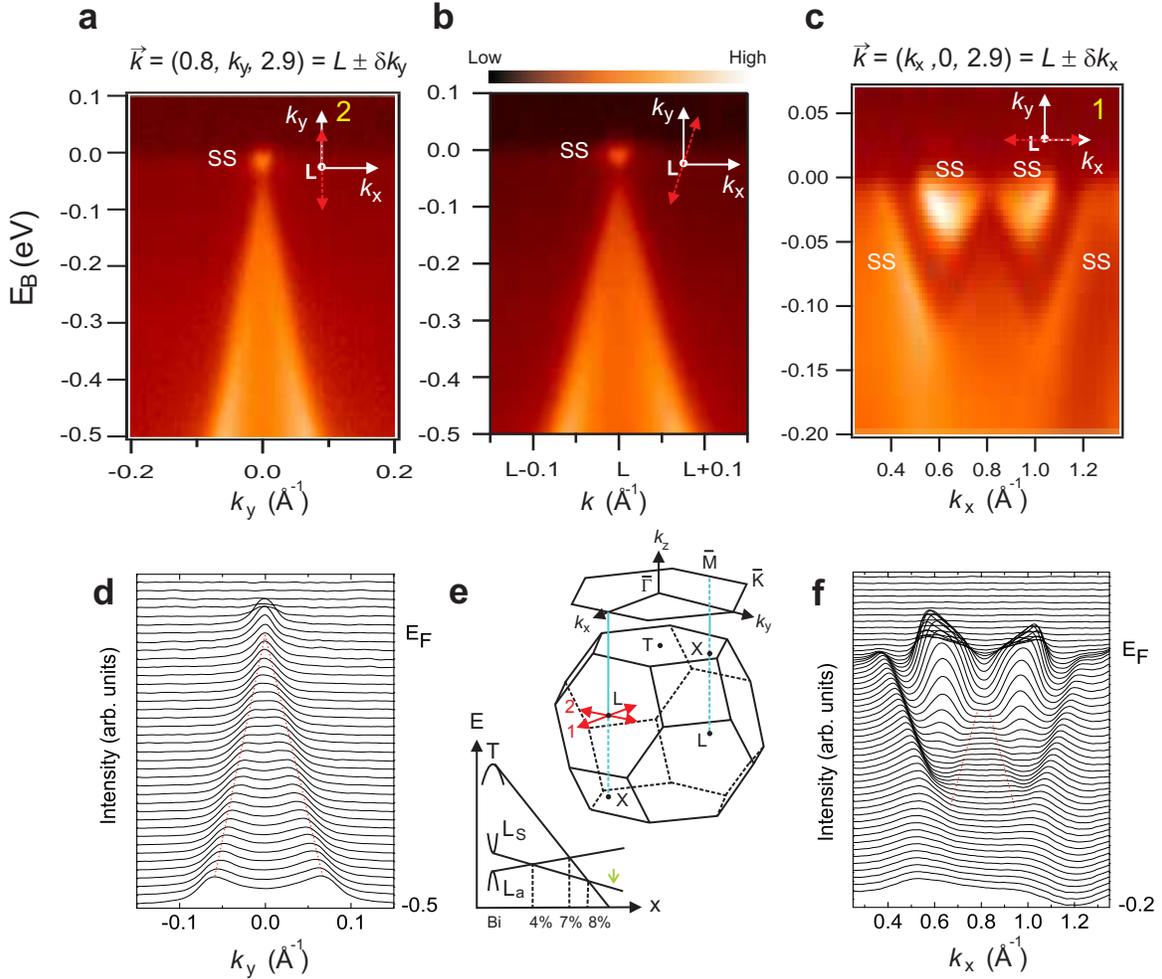}
\caption{\label{fig:BiSb_Fig2} \textbf{Dirac-like dispersion near
the L point in the bulk Brillouin zone.} Selected ARPES intensity
maps of Bi$_{0.9}$Sb$_{0.1}$ are shown along three $\vec{k}$-space
cuts through the L point of the bulk 3D Brillouin zone (BZ). The
presented data are taken in the third BZ with L$_z$ = 2.9 \AA$^{-1}$
with a photon energy of 29 eV. The cuts are along \textbf{a}, the
$k_y$ direction, \textbf{b}, a direction rotated by approximately
$10^{\circ}$ from the $k_y$ direction, and \textbf{c}, the $k_x$
direction. Each cut shows a $\Lambda$-shaped bulk band whose tip
lies below the Fermi level signalling a bulk gap. The surface states
are denoted SS and are all identified in Fig.2 (for further
identification via theoretical calculations see Supplementary
Information). \textbf{d}, Momentum distribution curves (MDCs)
corresponding to the intensity map in \textbf{a}. \textbf{f}, Log
scale plot of the MDCs corresponding to the intensity map in
\textbf{c}. The red lines are guides to the eye for the bulk
features in the MDCs. \textbf{e}, Schematic of the bulk 3D BZ of
Bi$_{1-x}$Sb$_x$ and the 2D BZ of the projected (111) surface. The
high symmetry points $\bar{\Gamma}$, $\bar{M}$ and $\bar{K}$ of the
surface BZ are labeled. Schematic evolution of bulk band energies as
a function of $x$ is shown. The L band inversion transition occurs
at $x \approx 0.04$, where a 3D gapless Dirac point is realized, and
the composition we study here (for which $x = 0.1$) is indicated by
the green arrow. A more detailed phase diagram based on our
experiments is shown in Fig.3c.}
\end{figure*}

Recently, semiconductors with inverted band gaps have been proposed
to manifest the two-dimensional (2D) quantum spin Hall phase, which
is predicted to be characterized by the presence of metallic 1D edge
states
\cite{Kane(Graphene),Bernevig(QSHE),Sheng(QSHE),Bernevig:HgTe}.
Although a band-inversion mechanism and edge states have been
invoked to interpret the transport results in 2D mercury telluride
semiconductor quantum wells \cite{Konig}, no 1D edge states are
directly imaged, so their topological character is unknown. Recent
theoretical treatments \cite{Fu:STI1, Murukami} have focused on the
strongly spin-orbit coupled, band-inverted Bi$_{1-x}$Sb$_x$ series
as a possible 3D bulk realization of the quantum spin Hall phase in
which the 1D edge states are expected to take the form of 2D surface
states \cite{Fu:STI1, Murukami, Fu:STI2} that may be directly imaged
and spectroscopically studied, making it feasible to identify their
topological order parameter character. Most importantly, the 3D
phase is a new phase of matter in terms of its topological
distinctions \cite{Moore:STI1}.

\begin{figure*}
\includegraphics[scale=0.65,clip=true, viewport=0.0in 0in 10.0in 7.4in]{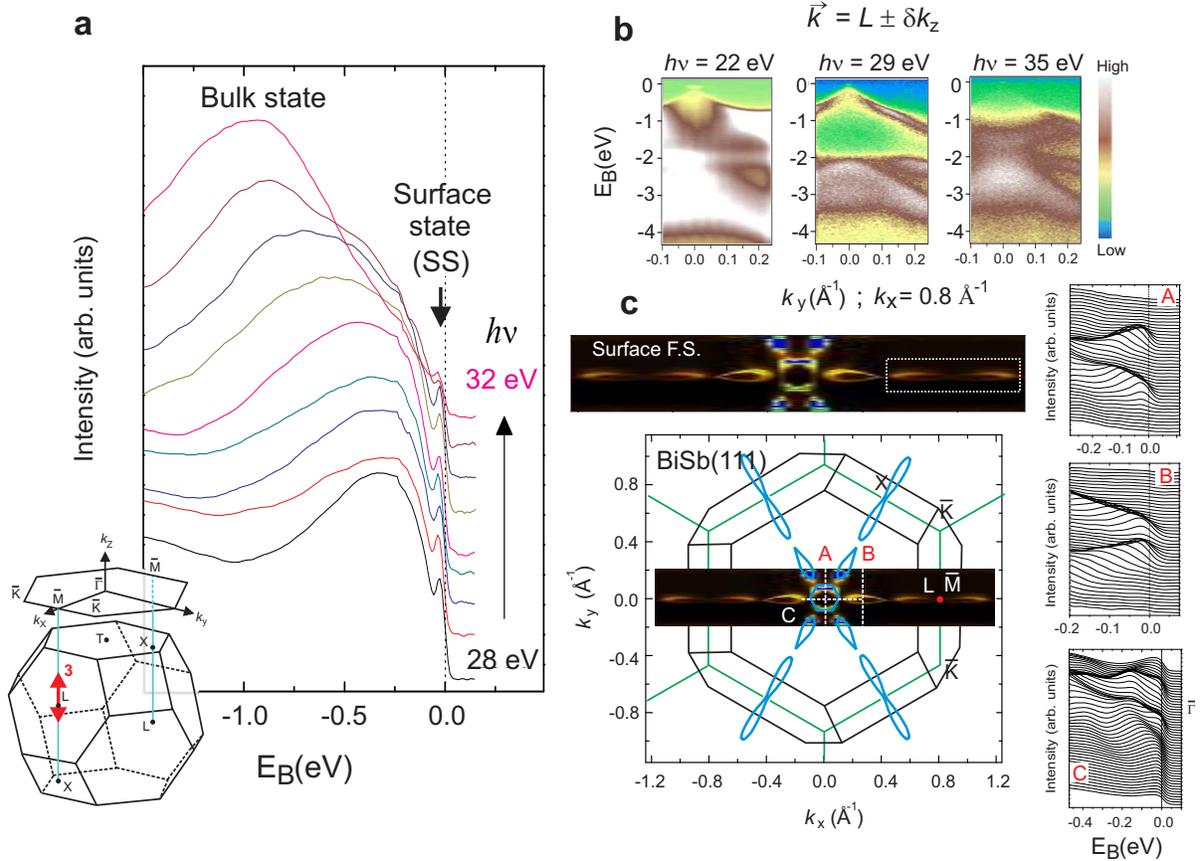}
\caption{\label{fig:BiSb_Fig2} \textbf{Dispersion along the cut in
the $\mathbf{k_z}$-direction.} Surface states are experimentally
identified by studying their out-of-plane momentum dispersion
through the systematic variation of incident photon energy.
\textbf{a}, Energy distribution curves (EDCs) of
Bi$_{0.9}$Sb$_{0.1}$ with electrons at the Fermi level ($E_F$)
maintained at a fixed in-plane momentum of ($k_x$=0.8 \AA$^{-1}$,
$k_y$=0.0 \AA$^{-1}$) are obtained as a function of incident photon
energy to identify states that exhibit no dispersion perpendicular
to the (111)-plane along the direction shown by the double-headed
arrow labeled ``3" in the inset (see Methods for detailed
procedure). Selected EDC data sets with photon energies of 28 eV to
32 eV in steps of 0.5 eV are shown for clarity. The non-energy
dispersive ($k_z$ independent) peaks near $E_F$ are the surface
states (SS). \textbf{b}, ARPES intensity maps along cuts parallel to
$k_y$ taken with electrons at $E_F$ fixed at $k_x$ = 0.8 \AA$^{-1}$
with respective photon energies of $h \nu$ = 22 eV, 29 eV and 35 eV
(for a conversion map from photon energy to $k_z$ see Supplementary
Information). The faint $\Lambda$-shaped band at $h \nu$ = 22 eV and
$h \nu$ = 35 eV shows some overlap with the bulk valence band at L
($h \nu$ = 29 eV), suggesting that it is a resonant surface state
degenerate with the bulk state in some limited k-range near $E_F$.
The flat band of intensity centered about $-$2 eV in the $h \nu$ =
22 eV scan originates from Bi 5d core level emission from second
order light. \textbf{c}, Projection of the bulk BZ (black lines)
onto the (111) surface BZ (green lines). Overlay (enlarged in inset)
shows the high resolution Fermi surface (FS) of the metallic SS
mode, which was obtained by integrating the ARPES intensity (taken
with $h \nu$ = 20 eV) from $-$15 meV to 10 meV relative to $E_F$.
The six tear-drop shaped lobes of the surface FS close to
$\bar{\Gamma}$ (center of BZ) show some intensity variation between
them that is due to the relative orientation between the axes of the
lobes and the axis of the detector slit. The six-fold symmetry was
however confirmed by rotating the sample in the $k_x-k_y$ plane.
EDCs corresponding to the cuts A, B and C are also shown; these
confirm the gapless character of the surface states in bulk
insulating Bi$_{0.9}$Sb$_{0.1}$.}
\end{figure*}

High-momentum-resolution angle-resolved photoemission spectroscopy
performed with varying incident photon energy (IPEM-ARPES) allows
for measurement of electronic band dispersion along various momentum
space ($\vec{k}$-space) trajectories in the 3D bulk BZ. ARPES
spectra taken along two orthogonal cuts through the L point of the
bulk BZ of Bi$_{0.9}$Sb$_{0.1}$ are shown in Figs 1a and c. A
$\Lambda$-shaped dispersion whose tip lies less than 50 meV below
the Fermi energy ($E_F$) can be seen along both directions.
Additional features originating from surface states that do not
disperse with incident photon energy are also seen. Owing to the
finite intensity between the bulk and surface states, the exact
binding energy ($E_B$) where the tip of the $\Lambda$-shaped band
dispersion lies is unresolved. The linearity of the bulk
$\Lambda$-shaped bands is observed by locating the peak positions at
higher $E_B$ in the momentum distribution curves (MDCs), and the
energy at which these peaks merge is obtained by extrapolating
linear fits to the MDCs. Therefore 50 meV represents a lower bound
on the energy gap $\Delta$ between L$_a$ and L$_s$. The magnitude of
the extracted band velocities along the $k_x$ and $k_y$ directions
are $7.9 \pm 0.5 \times 10^4$ ms$^{-1}$ and $10.0 \pm 0.5 \times
10^5$ ms$^{-1}$, respectively, which are similar to the tight
binding values $7.6 \times 10^4$ ms$^{-1}$ and $9.1 \times 10^5$
ms$^{-1}$ calculated for the L$_a$ band of bismuth \cite{Liu}. Our
data are consistent with the extremely small effective mass of
$0.002m_e$ (where $m_e$ is the electron mass) observed in
magneto-reflection measurements on samples with $x = 11\%$
\cite{Hebel}. The Dirac point in graphene, co-incidentally, has a
band velocity ($|v_F| \approx 10^6$ ms$^{-1}$) \cite{Zhang}
comparable to what we observe for Bi$_{0.9}$Sb$_{0.1}$, but its
spin-orbit coupling is several orders of magnitude weaker
\cite{Kane(Graphene)}, and the only known method of inducing a gap
in the Dirac spectrum of graphene is by coupling to an external
chemical substrate \cite{Zhou}. The Bi$_{1-x}$Sb$_x$ series thus
provides a rare opportunity to study relativistic Dirac Hamiltonian
physics in a 3D condensed matter system where the intrinsic (rest)
mass gap can be easily tuned.

\begin{figure*}
\includegraphics[scale=0.65,clip=true, viewport=0.0in 0in 10.5in 6.7in]{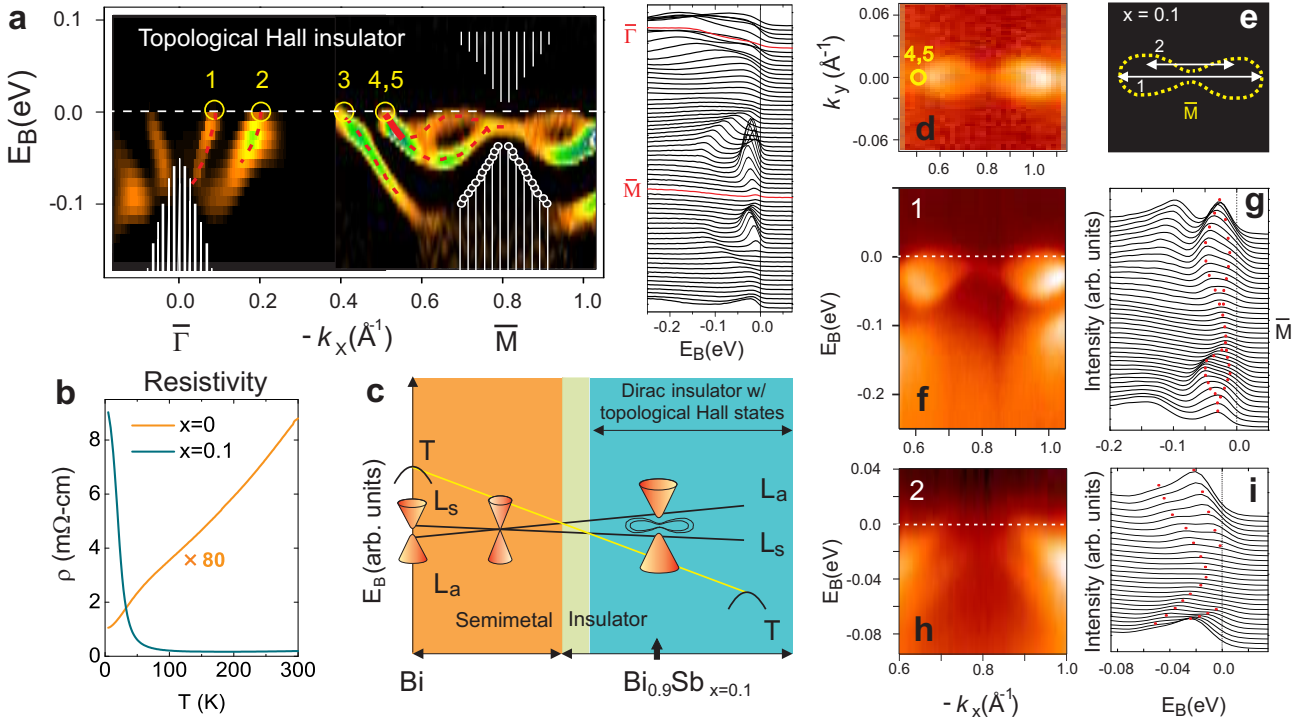}
\caption{\label{fig:BiSb_Fig2} \textbf{The topological gapless
surface states in bulk insulating Bi$_{0.9}$Sb$_{0.1}$.} \textbf{a},
The surface band dispersion second derivative image (SDI) of
Bi$_{0.9}$Sb$_{0.1}$ along $\bar{\Gamma} - \bar{M}$. The shaded
white area shows the projection of the bulk bands based on ARPES
data, as well as a rigid shift of the tight binding bands to sketch
the unoccupied bands above the Fermi level. To maintain high
momentum resolution, data were collected in two segments of momentum
space, then the intensities were normalized using background level
above the Fermi level. A non-intrinsic flat band of intensity near
$E_F$ generated by the SDI analysis was rejected to isolate the
intrinsic dispersion. The Fermi crossings of the surface state are
denoted by yellow circles, with the band near $-k_x \approx 0.5$
\AA$^{-1}$ counted twice owing to double degeneracy. The red lines
are guides to the eye. An in-plane rotation of the sample by
$60^{\circ}$ produced the same surface state dispersion. The EDCs
along $\bar{\Gamma} - \bar{M}$ are shown to the right. There are a
total of five crossings from $\bar{\Gamma} - \bar{M}$ which
indicates that these surface states are topologically non-trivial.
The number of surface state crossings in a material (with an odd
number of Dirac points) is related to the topological $Z_2$
invariant (see text). \textbf{b}, The resistivity curves of Bi and
Bi$_{0.9}$Sb$_{0.1}$ reflect the contrasting transport behaviours.
The presented resistivity curve for pure bismuth has been multiplied
by a factor of 80 for clarity. \textbf{c}, Schematic variation of
bulk band energies of Bi$_{1-x}$Sb$_x$ as a function of $x$ (based
on band calculations and on \cite{Fu:STI1, Lenoir}).
Bi$_{0.9}$Sb$_{0.1}$ is a direct gap bulk Dirac point insulator well
inside the inverted-band regime, and its surface forms a
``topological metal'' - the 2D analogue of the 1D edge states in
quantum spin Hall systems. \textbf{d}, ARPES intensity integrated
within $\pm 10$ meV of $E_F$ originating solely from the surface
state crossings. The image was plotted by stacking along the
negative $k_x$ direction a series of scans taken parallel to the
$k_y$ direction. \textbf{e}, Outline of Bi$_{0.9}$Sb$_{0.1}$ surface
state ARPES intensity near $E_F$ measured in \textbf{d}. White lines
show scan directions ``1'' and ``2''. \textbf{f}, Surface band
dispersion along direction ``1'' taken with $h \nu$ = 28 eV and the
corresponding EDCs (\textbf{g}). The surface Kramers degenerate
point, critical in determining the topological $Z_2$ class of a band
insulator, is clearly seen at $\bar{M}$, approximately $15 \pm 5$
meV below $E_F$. (We note that the scans are taken along the
negative $k_x$ direction, away from the bulk L point.) \textbf{h},
Surface band dispersion along direction ``2'' taken with $h \nu$
 = 28 eV and the corresponding EDCs (\textbf{i}). This scan no longer
passes through the $\bar{M}$-point, and the observation of two well
separated bands indicates the absence of Kramers degeneracy as
expected, which cross-checks the result in (\textbf{a}).}
\end{figure*}

Studying the band dispersion perpendicular to the sample surface
provides a way to differentiate bulk states from surface states in a
3D material. To visualize the near-$E_F$ dispersion along the 3D L-X
cut (X is a point that is displaced from L by a $k_z$ distance of
3$\pi/c$, where $c$ is the lattice constant), in Fig.2a we plot
energy distribution curves (EDCs), taken such that electrons at
$E_F$ have fixed in-plane momentum $(k_x, k_y)$ = (L$_x$, L$_y$) =
(0.8 \AA$^{-1}$, 0.0 \AA$^{-1}$), as a function of photon energy
($h\nu$). There are three prominent features in the EDCs: a
non-dispersing, $k_z$ independent, peak centered just below $E_F$ at
about $-$0.02 eV; a broad non-dispersing hump centered near $-$0.3
eV; and a strongly dispersing hump that coincides with the latter
near $h\nu$ = 29 eV. To understand which bands these features
originate from, we show ARPES intensity maps along an in-plane cut
$\bar{K} \bar{M} \bar{K}$ (parallel to the $k_y$ direction) taken
using $h\nu$ values of 22 eV, 29 eV and 35 eV, which correspond to
approximate $k_z$ values of L$_z -$ 0.3 \AA$^{-1}$, L$_z$, and L$_z$
+ 0.3 \AA$^{-1}$ respectively (Fig.2b). At $h\nu$ = 29 eV, the low
energy ARPES spectral weight reveals a clear $\Lambda$-shaped band
close to $E_F$. As the photon energy is either increased or
decreased from 29 eV, this intensity shifts to higher binding
energies as the spectral weight evolves from the $\Lambda$-shaped
into a $\cup$-shaped band. Therefore the dispersive peak in Fig.2a
comes from the bulk valence band, and for $h\nu$ = 29 eV the high
symmetry point L = (0.8, 0, 2.9) appears in the third bulk BZ. In
the maps of Fig.2b with respective $h\nu$ values of 22 eV and 35 eV,
overall weak features near $E_F$ that vary in intensity remain even
as the bulk valence band moves far below $E_F$. The survival of
these weak features over a large photon energy range (17 to 55 eV)
supports their surface origin. The non-dispersing feature centered
near $-0.3$ eV in Fig.2a comes from the higher binding energy
(valence band) part of the full spectrum of surface states, and the
weak non-dispersing peak at $-0.02$ eV reflects the low energy part
of the surface states that cross $E_F$ away from the $\bar{M}$ point
and forms the surface Fermi surface (Fig.2c).

Having established the existence of an energy gap in the bulk state
of Bi$_{0.9}$Sb$_{0.1}$ (Figs 1 and 2) and observed linearly
dispersive bulk bands uniquely consistent with strong spin-orbit
coupling model calculations \cite{Wolff, Fukuyama, Buot, Liu} (see
Supplementary Information for full comparison with theoretical
calculation), we now discuss the topological character of its
surface states, which are found to be gapless (Fig.2c). In general,
the states at the surface of spin-orbit coupled compounds are
allowed to be spin split owing to the loss of space inversion
symmetry $[E(k,\uparrow) = E(-k,\uparrow)]$. However, as required by
Kramers' theorem, this splitting must go to zero at the four time
reversal invariant momenta (TRIM) in the 2D surface BZ. As discussed
in \cite{Fu:STI1, Fu:STI2}, along a path connecting two TRIM in the
same BZ, the Fermi energy inside the bulk gap will intersect these
singly degenerate surface states either an even or odd number of
times. When there are an even number of surface state crossings, the
surface states are topologically trivial because weak disorder (as
may arise through alloying) or correlations can remove \emph{pairs}
of such crossings by pushing the surface bands entirely above or
below $E_F$. When there are an odd number of crossings, however, at
least one surface state must remain gapless, which makes it
non-trivial \cite{Fu:STI1, Murukami, Fu:STI2}. The existence of such
topologically non-trivial surface states can be theoretically
predicted on the basis of the \emph{bulk} band structure only, using
the $Z_2$ invariant that is related to the quantum Hall Chern number
\cite{Kane(QSHE-Z2)}. Materials with band structures with $Z_2 = +1$
are ordinary Bloch band insulators that are topologically equivalent
to the filled shell atomic insulator, and are predicted to exhibit
an even number (including zero) of surface state crossings.
Materials with bulk band structures with $Z_2 = -1$ on the other
hand, which are expected to exist in rare systems with strong
spin-orbit coupling acting as an internal quantizing magnetic field
on the electron system \cite{Haldane(P-anomaly)}, and inverted bands
at an odd number of high symmetry points in their bulk 3D BZs, are
predicted to exhibit an odd number of surface state crossings,
precluding their adiabatic continuation to the atomic insulator
\cite{Kane(Graphene), Fu:STI1, Murukami, Fu:STI2, Moore:STI1, Roy,
Bernevig:HgTe, Konig}. Such ``topological quantum Hall metals''
\cite{Fu:STI2, Moore:STI1, Roy} cannot be realized in a purely 2D
electron gas system such as the one realized at the interface of
GaAs/GaAlAs systems.

In our experimental case, namely the (111) surface of
Bi$_{0.9}$Sb$_{0.1}$, the four TRIM are located at $\bar{\Gamma}$
and three $\bar{M}$ points that are rotated by $60^{\circ}$ relative
to one another. Owing to the three-fold crystal symmetry (A7 bulk
structure) and the observed mirror symmetry of the surface Fermi
surface across $k_x = 0$ (Fig.2), these three $\bar{M}$ points are
equivalent (and we henceforth refer to them as a single point,
$\bar{M}$). The mirror symmetry $[E(k_y) = E(-k_y)]$ is also
expected from time reversal invariance exhibited by the system. The
complete details of the surface state dispersion observed in our
experiments along a path connecting $\bar{\Gamma}$ and $\bar{M}$ are
shown in Fig.3a; finding this information is made possible by our
experimental separation of surface states from bulk states. As for
bismuth (Bi), two surface bands emerge from the bulk band continuum
near $\bar{\Gamma}$ to form a central electron pocket and an
adjacent hole lobe \cite{Ast:Bi1, Hochst,Hofmann}. It has been
established that these two bands result from the spin-splitting of a
surface state and are thus singly degenerate \cite{Hirahara,
Hofmann}. On the other hand, the surface band that crosses $E_F$ at
$-k_x \approx 0.5$ \AA$^{-1}$, and forms the narrow electron pocket
around $\bar{M}$, is clearly doubly degenerate, as far as we can
determine within our experimental resolution. This is indicated by
its splitting below $E_F$ between $-k_x \approx 0.55$ \AA$^{-1}$ and
$\bar{M}$, as well as the fact that this splitting goes to zero at
$\bar{M}$ in accordance with Kramers theorem. In semimetallic single
crystal bismuth, only a single surface band is observed to form the
electron pocket around $\bar{M}$ \cite{Hengsberger, Ast:Bi2}.
Moreover, this surface state overlaps, hence becomes degenerate
with, the bulk conduction band at L (L projects to the surface
$\bar{M}$ point) owing to the semimetallic character of Bi (Fig.3b).
In Bi$_{0.9}$Sb$_{0.1}$ on the other hand, the states near $\bar{M}$
fall completely inside the bulk energy gap preserving their purely
surface character at $\bar{M}$ (Fig.3a). The surface Kramers doublet
point can thus be defined in the bulk insulator (unlike in Bi
\cite{Hirahara,Ast:Bi1, Hochst, Hofmann, Hengsberger, Ast:Bi2}) and
is experimentally located in Bi$_{0.9}$Sb$_{0.1}$ samples to lie
approximately 15 $\pm$ 5 meV below $E_F$ at $\vec{k} = \bar{M}$
(Fig.3a). For the precise location of this Kramers point, it is
important to demonstrate that our alignment is strictly along the
$\bar{\Gamma} - \bar{M}$ line. To do so, we contrast high resolution
ARPES measurements taken along the $\bar{\Gamma} - \bar{M}$ line
with those that are slightly offset from it (Fig.3e). Figs 3f-i show
that with $k_y$ offset from the Kramers point at $\bar{M}$ by less
than 0.02 \AA$^{-1}$, the degeneracy is lifted and only one band
crosses $E_F$ to form part of the bow-shaped electron distribution
(Fig.3d). Our finding of five surface state crossings (an odd rather
than an even number) between $\bar{\Gamma}$ and $\bar{M}$ (Fig.3a),
confirmed by our observation of the Kramers degenerate point at the
TRIM, indicates that these gapless surface states are topologically
non-trivial. This corroborates our bulk electronic structure result
that Bi$_{0.9}$Sb$_{0.1}$ is in the insulating band-inverted ($Z_2 =
-1$) regime (Fig.3c), which contains an odd number of bulk (gapped)
Dirac points, and is topologically analogous to an integer quantum
spin Hall insulator.

\setcounter{figure}{0}

\begin{figure*}
\renewcommand{\thefigure}{S\arabic{figure}}
\includegraphics[scale=0.65,clip=true, viewport=0.0in 0in 8.0in 6.3in]{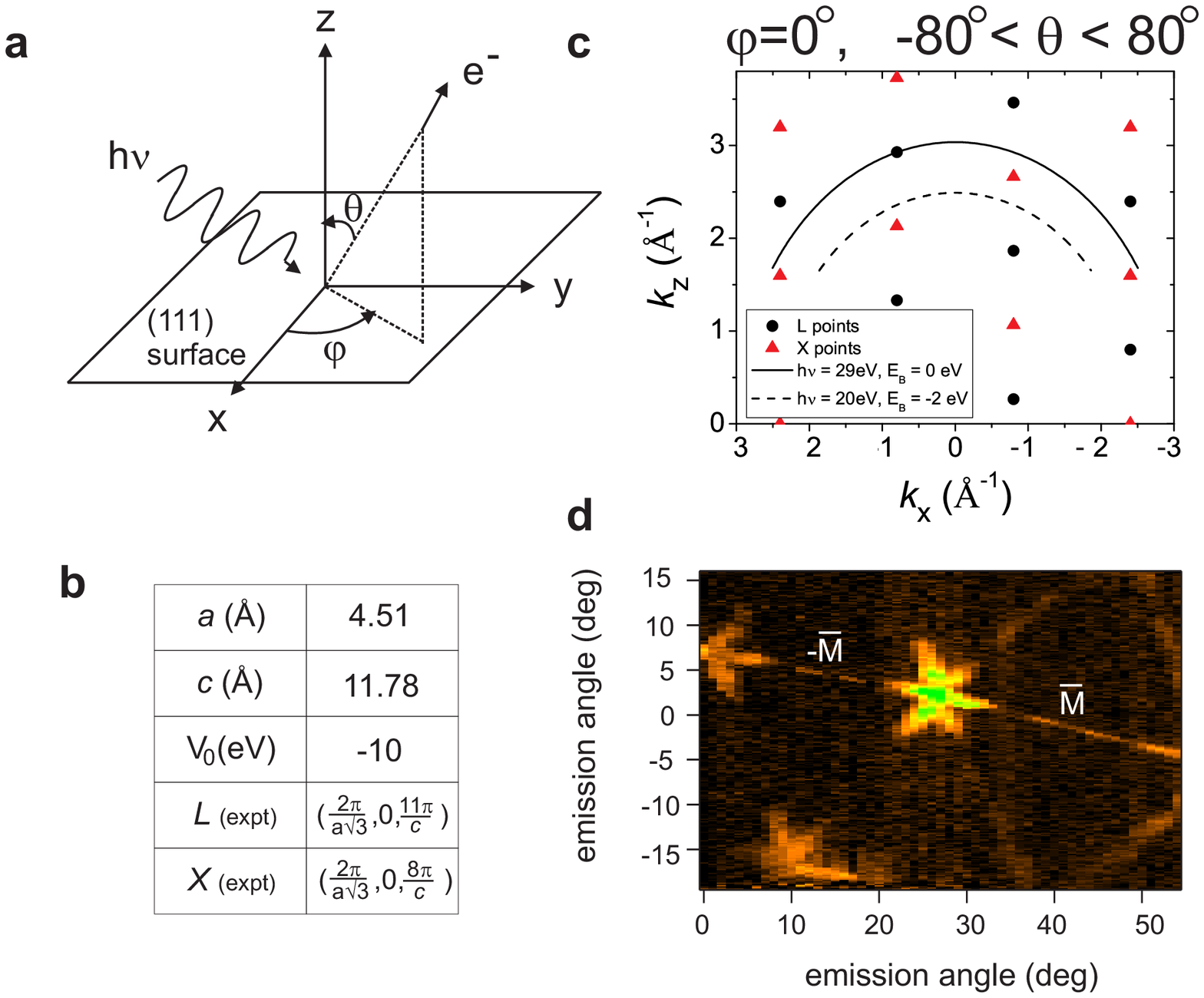}
\caption{\textbf{Method of locating high symmetry bulk reciprocal
lattice points of Bi$_{0.9}$Sb$_{0.1}$ using incident photon energy
modulated ARPES.} \textbf{a,} Geometry of an ARPES experiment.
\textbf{b,} Key parameters relevant to the calculation of the
positions of the high symmetry points in the 3D BZ. The lattice
constants refer to the rhombohedral A7 lattice structure.
\textbf{c,} Location of $L$ and $X$ points of the bulk BZ in the
$k_x$-$k_z$ plane together with the constant energy contours that
can be accessed by changing the angle $\theta$. \textbf{d,} Near
$E_F$ intensity map ($h\nu$ = 55 eV) of the Fermi surface formed by
the surface states covering an entire surface BZ, used to help
locate various in-plane momenta, in units of the photoelectron
emission angle along two orthogonal spatial directions. The electron
pockets near \={M} in Fig.2c (main text) appear as lines in Fig.S1d
due to relaxed k-resolution in order to cover a large k-space in a
single shot.}
\end{figure*}

Our experimental results taken collectively strongly suggest that
Bi$_{0.9}$Sb$_{0.1}$ is quite distinct from graphene \cite{Zhang,
Novoselov} and represents a novel state of quantum matter: a
strongly spin-orbit coupled insulator with an odd number of Dirac
points with a negative $Z_2$ topological Hall phase, which realizes
the ``parity anomaly without Fermion doubling". Our work further
demonstrates a general methodology for possible future
investigations of \emph{novel topological orders} in exotic quantum
matter.

\textbf{Note Added :
In a very recent work we have successfully imaged the spin-polarization of the topological edge modes using high-resolution spin-resolved-ARPES \cite{Science}.}

\vspace{0.5cm}

\small{\textbf{Acknowledgements} We thank P. W. Anderson, B. A.
Bernevig, F. D. M. Haldane, D. A. Huse, C. L. Kane, R. B. Laughlin,
N. P. Ong, A. N. Pasupathy and D. C. Tsui for discussions. This work is supported by the DOE Office of Basic Energy Science and materials synthesis is supported by the NSF MRSEC.}

\vspace{0.5cm}

\small{\textbf{Author information} Correspondence and requests for
materials should be addressed to M.Z.H (mzhasan@princeton.edu).}


\section{Methods Summary}

High resolution IPEM-ARPES data have been taken at Beamlines 12.0.1
and 10.0.1 of the Advanced Light Source in Lawrence Berkeley
National Laboratory, as well as at PGM Beamline of the Synchrotron
Radiation Center in Wisconsin, with photon energies from 17 to 55 eV
and energy resolution from 9 to 40 meV and momentum (k-)resolution
better than $1.5 \%$ of the surface Brillouin zone. Data were taken
on high quality bulk single crystal Bi$_{1-x}$Sb$_x$ at a
temperature of 15 K and chamber pressures better than $8 \times
10^{-11}$ torr. Throughout this paper, the bulk bands presented are
from those measured in the third bulk Brillouin zone to ensure a
high degree of signal-to-noise contrast, and the $k_z$ values are
estimated using the standard free-electron final state
approximation.

\begin{figure*}
\renewcommand{\thefigure}{S\arabic{figure}}
\includegraphics[scale=0.75,clip=true, viewport=-0.15in 0in 11.0in 6.0in]{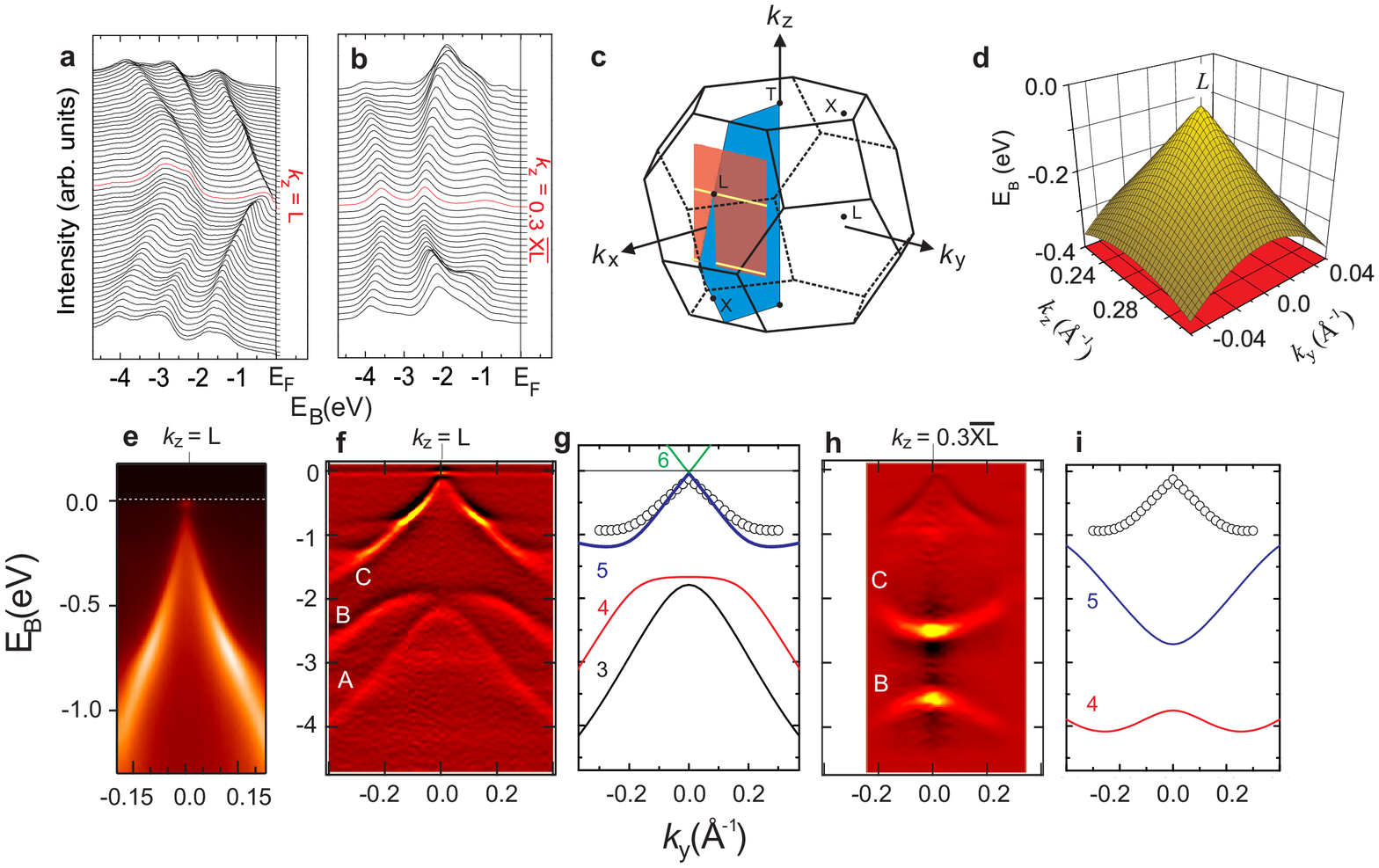}
\caption{\textbf{Identification of the bulk band features of
Bi$_{0.9}$Sb$_{0.1}$. Experimental band-structure determined by
ARPES is compared to bulk tight-binding calculations of bismuth to
further identify the deeper lying bulk bands and their symmetry
origins.} \textbf{a,} Energy distribution curves (EDCs) along a
k-space cut given by the upper yellow line shown in schematic
\textbf{c} which goes through the bulk $L$ point in the 3$^{rd}$ BZ
($h\nu$ = 29 eV). The corresponding ARPES intensity in the vicinity
of $L$ is shown in \textbf{e}. \textbf{b,} EDCs along the lower
yellow line of \textbf{c} which goes through the point a fraction
0.3 of the k-distance from $X$ to $L$ ($h\nu$ = 20 eV), showing a
dramatic change of the deeper lying band dispersions. (This cut was
taken at a $k_x$ value equal in magnitude but opposite in sign to
that in \textbf{a} as described in the SI text). \textbf{f,h,} The
ARPES second derivative images (SDI) of the raw data shown in
\textbf{a} and \textbf{b} to reveal the band dispersions. The flat
band of intensity at $E_F$ is an artifact of taking SDI.
\textbf{g,i,} Tight binding band calculations of bismuth including
spin-orbit coupling, using Liu and Allen model$^{22}$, along the
corresponding experimental cut directions shown in \textbf{f} and
\textbf{h}. The bands (colored solid lines) labelled 3 to 6 are
derived from the symmetries associated with the 6$p$-orbitals and
their dispersion is thus strongly influenced by spin-orbit coupling.
The inter-band gap between bands 5 and 6 is barely visible on the
scale of Fig. S2g. The circled curves mark the surface state
dispersion, which is present at all measured photon energies (no
$k_z$ dispersion). There is a close match of the bulk band
dispersion between the data and calculations, confirming the
presence of strong spin-orbit coupling. \textbf{d,} Tight binding
valence band (5) dispersion of bismuth in the $k_y$-$k_z$ momentum
plane showing linearity along both directions. The close match
between data and calculation along $k_y$ suggests that the
dispersion near $E_F$ along $k_z$ is also linear.}
\end{figure*}

\section{SUPPLEMENTARY INFORMATION}

\section{METHODS}

\subsection{Growth method for high-quality single crystals}

\begin{figure*}
\renewcommand{\thefigure}{S\arabic{figure}}
\includegraphics[scale=0.55,clip=true, viewport=0.0in 0in 10.0in 8.5in]{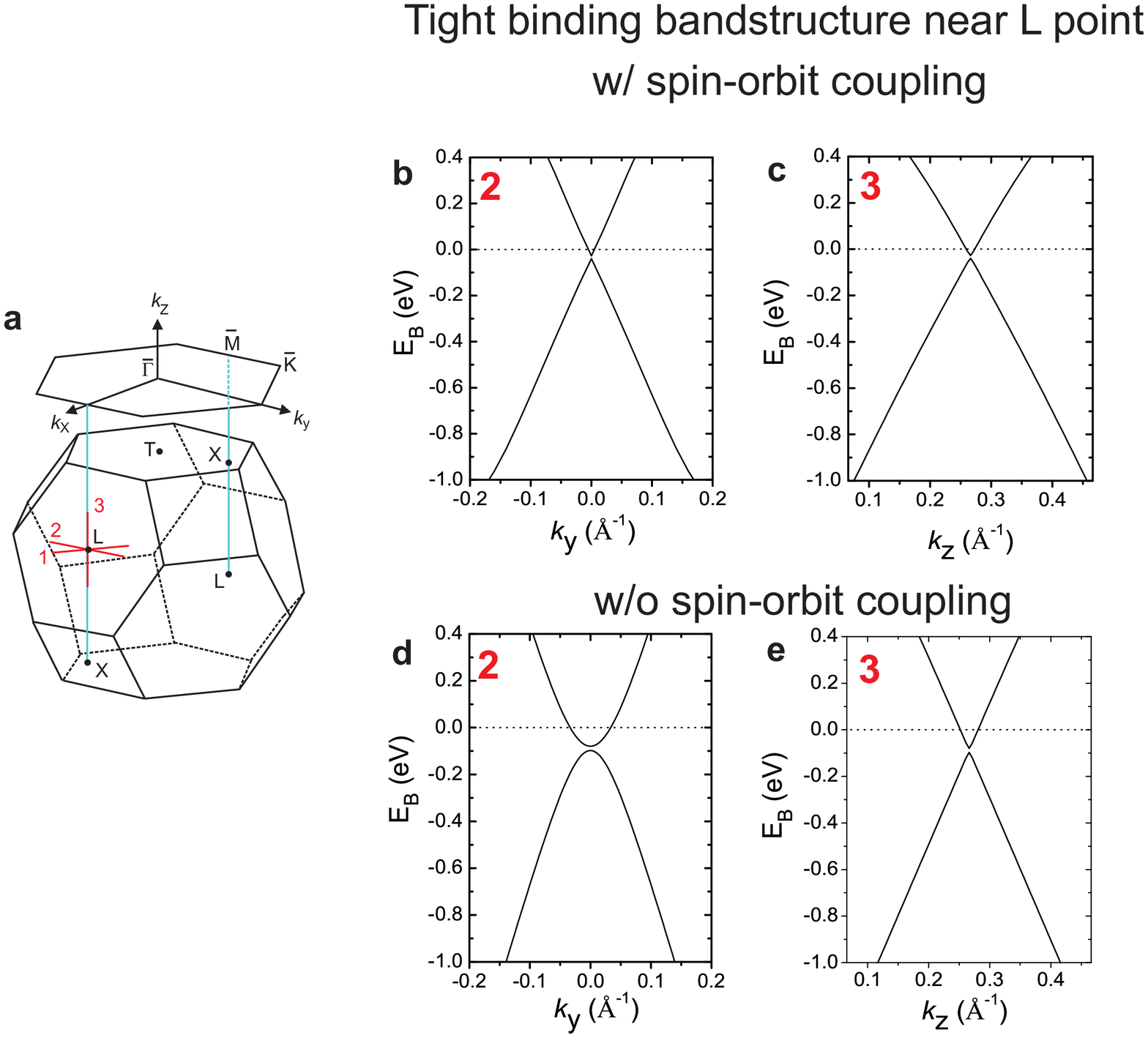}
\caption{\textbf{Spin-orbit coupling has a profound effect on the
band structure of bismuth near $L$ point.} \textbf{a,} Schematic of
the bulk 3D BZ and the projected BZ of the (111) surface.
\textbf{b,c,} Calculated tight binding band structure of bismuth
including a spin-orbit coupling strength of 1.5 eV $^{22}$ along two
orthogonal cuts through the $L$ point in the 1$^{st}$ bulk BZ.
\textbf{d,e,} Tight binding band structure along the same two
directions as \textbf{b} and \textbf{c} calculated without
spin-orbit coupling. The inter-band gap of 13.7 meV is barely
visible on the scale of \textbf{b} and \textbf{c}.}
\end{figure*}

The Bi$_{1-x}$Sb$_x$ single-crystal samples ($0\leq x \leq0.17$)
used for ARPES experiments were each cleaved from a boule grown from
a stoichiometric mixture of high-purity elements. The boule was
cooled from 650 $^{\circ}$C to 270 $^{\circ}$C over a period of five
days and was annealed for seven days at 270 $^{\circ}$C. The samples
naturally cleaved along the (111) plane, which resulted in shiny
flat silver surfaces. X-ray diffraction measurements were used to
check that the samples were single phase, and confirmed that the
Bi$_{0.9}$Sb$_{0.1}$ single crystals presented in this paper have a
rhombohedral A7 crystal structure (point group $R\bar{3}m$), with
room-temperature (T=300K) lattice parameters $a$ = 4.51 \AA\ and $c$
= 11.78 \AA\ indexed using a rhombohedral unit cell. The X-ray
diffraction patterns of the cleaved crystals exhibit only the (333),
(666), and (999) peaks, showing that the cleaved surface is oriented
along the trigonal (111) axis. Room-temperature data were recorded
on a Bruker D8 diffractometer using Cu K$\alpha$  radiation
($\lambda$ = 1.54 \AA) and a diffracted-beam monochromator. The
in-plane crystal orientation was determined by Laue X-ray
diffraction. During the angle-resolved photoemission spectroscopy
(ARPES) measurements a fine alignment was achieved by carefully
studying the band dispersions and Fermi surface symmetry as an
internal check for crystal orientation.

\subsection{Transport measurements}

Temperature-dependent resistivity measurements were carried out on
single-crystal samples in a Quantum Design PPMS-9 instrument, using
a standard four-probe technique on approximately 4 $\times$ 1
$\times$1-mm$^3$, rectangular samples with the current in the basal
plane, which was perpendicular to the trigonal axis. The four
contacts were made by using room-temperature silver paste. The data
for samples with concentrations ranging from $x$ = 0 to $x$ = 0.17
showed a systematic change from semimetallic to insulating-like
behaviour with increasing $x$, in agreement with previously
published works$^{15}$, which was used as a further check of the
antimony concentrations. Conventional magnetic and transport
measurements$^{7,17,31}$ such as these cannot separately measure the
contributions of the surface and bulk states to the total signal.
ARPES, on the other hand, is a momentum-selective technique$^{32}$,
which allows for a separation of 2D (surface) from 3D (bulk)
dispersive energy bands. This capability is especially important for
Bi$_{1-x}$Sb$_x$ because the Dirac point lies at a single point in
the 3D Brillouin zone, unlike for 2D graphene, where the Dirac
points can be studied at any arbitrary perpendicular momentum along
a line$^{33,34}$.

\subsection{Systematic methods for separating bulk from surface
electronic states}

ARPES is a photon-in, electron-out technique$^{32}$. Photoelectrons
ejected from a sample by a monochromatic beam of radiation are
collected by a detector capable of measuring its kinetic energy
$E_{kin}$. By varying the detector angles, $\theta$ and $\varphi$,
relative to the sample surface normal, the momentum of the
photoelectrons, $\textbf{K}$, can also be determined (as illustrated
in Supplementary Fig. 1a). By employing the commonly used
free-electron final state approximation, we can fully convert from
the measured kinetic energy and momentum values of the photoelectron
to the binding energy, $E_B$, and Bloch momentum values $\textbf{k}$
of its initial state inside the crystal, via

\begin{align}
|E_B| &= h\nu - W - E_{kin}\notag\\
k_x &= K_x = \frac{1}{\hbar}\sqrt{2 m_e
E_{kin}}\textnormal{sin}\theta\notag\\
k_z &=
\frac{1}{\hbar}\sqrt{2m_e(E_{kin}\textnormal{cos}^2\theta-V_0)}\notag
\end{align}
where we have set $\varphi$ = 0, $W$ is the work function, $m_e$ is
the electron mass and $V_0$ is an experimentally determined
parameter, which is approximately $-$10 eV for bismuth$^{35,36}$.
Features in the ARPES spectra originating from bulk initial states
(dispersive along the $k_z$-direction) were distinguished from those
originating from surface initial states (non-dispersive along the
$k_z$-direction) by studying their dependence on incident photon
energy, $h\nu$, and converting this to dependence on $k_z$ via the
displayed equations. ARPES data were collected at beamlines 12.0.1
and 10.0.1 of the Advanced Light Source at the Lawrence Berkeley
National Laboratory, as well as at the PGM beamline of the
Synchrotron Radiation Center in Wisconsin, with incident photon
energies ranging from 17 eV to 55 eV, energy resolutions ranging
from 9 meV to 40 meV and momentum resolution better than 1.5\% of
the surface Brillouin zone, using Scienta electron analysers. The
combination of high spatial resolution and high crystalline quality
enabled us to probe only the highly ordered and cleanest regions of
our samples. Single-crystal Bi$_{1-x}$Sb$_x$ samples were cleaved in
situ at a temperature of 15 K and chamber pressures less than 8
$\times$ 10$^{-11}$ torr, and high surface quality was checked
throughout the measurement process by monitoring the EDC linewidths
of the surface state. To measure the near-$E_F$ dispersion of an
electronic band along a direction normal to the sample surface, such
as the direction from $X (2\pi/\sqrt{3}a, 0, 8\pi/c)$ to $L
(2\pi/\sqrt{3}a, 0, 11\pi/c)$ shown in Fig. 2a, EDCs were taken at
several incident photon energies. The kinetic energy of the
photoelectron at $E_F$ is different for each value of $h\nu$, so the
angle was first adjusted and then held fixed for each $h\nu$ so as
to keep $k_x$ constant at $2\pi/\sqrt{3}a$ = 0.8 \AA$^{-1}$ for
electrons emitted near $E_F$. To ensure that the in-plane momentum
remained constant at \={M} (the $L$-$X$ line projects onto \={M})
for each EDC, a complete near-$E_F$ intensity map was generated for
each photon energy to precisely locate the \={M}-point (see
Supplementary Fig. 1d). We note that because the bulk crystal has
only three-fold rotational symmetry about the $k_z$-axis, the
reciprocal lattice does not have mirror symmetry about the $k_x$ = 0
plane. Therefore, scans taken at +$\theta$ and -$\theta$  for the
same photon energy probe different points in the bulk 3D Brillouin
zone; this is responsible for the absence of the bulk
$\Lambda$-shaped band in Fig. 3f.

\section{Confirming the bulk nature of electronic bands by
comparison with theoretical calculations}

In an ARPES experiment (Fig.S1a), three dimensional (3D) dispersive
bulk electronic states can be identified as those that disperse with
incident photon energy, whereas surface states do not. As an
additional check that we have indeed correctly identified the bulk
bands of Bi$_{0.9}$Sb$_{0.1}$ in Figs 1 and 2, we also measured the
dispersion of the deeper lying bands well below the Fermi level
($E_F$) and compared them to tight binding theoretical calculations
of the bulk bands of pure bismuth following the model of Liu and
Allen (1995)$^{22}$. A tight-binding approach is known to be valid
since Bi$_{0.9}$Sb$_{0.1}$ is not a strongly correlated electron
system. As Bi$_{0.9}$Sb$_{0.1}$ is a random alloy (Sb does not form
a superlattice$^{17}$) with a relatively small Sb concentration
($\sim$0.2 Sb atoms per rhombohedral unit cell), the deeper lying
band structure of Bi$_{0.9}$Sb$_{0.1}$ is expected to follow that of
pure Bi because the deeper lying (localized wave function) bands of
Bi$_{0.9}$Sb$_{0.1}$ are not greatly affected by the substitutional
disorder, and no additional back folded bands are expected to arise.
Since these deeper lying bands are predicted to change dramatically
with $k_z$, they help us to finely determine the experimentally
probed $k_z$ values. Fig.S2f shows the ARPES second derivative image
(SDI) of a cut parallel to \={K}\={M}\={K} that passes through the
$L$ point of the 3D Brillouin zone (BZ), and Fig.S2h shows a
parallel cut that passes through the 0.3 $XL$ point (Fig.S2c). The
locations of these two cuts in the 3D bulk BZ were calculated from
the kinematic relations described in the Methods section, from which
we can construct the constant energy contours shown in Fig. S1c. By
adjusting $\theta$ such that the in-plane momentum $k_x$ is fixed at
approximately 0.8 \AA$^{-1}$ (the surface \={M} point), at a photon
energy $h\nu$ =29 eV, electrons at the Fermi energy ($E_B$=0 eV)
have a $k_z$ that corresponds to the $L$ point in the 3$^{rd}$ bulk
BZ. By adjusting $\theta$ such that the in-plane momentum $k_x$ is
fixed at approximately -0.8 \AA$^{-1}$, at a photon energy $h\nu$ =
20 eV, electrons at a binding energy of -2 eV have a $k_z$ near 0.3
$XL$.

\begin{figure*}
\renewcommand{\thefigure}{S\arabic{figure}}
\includegraphics[scale=0.6,clip=true, viewport=0.0in 0in 10.0in 8.0in]{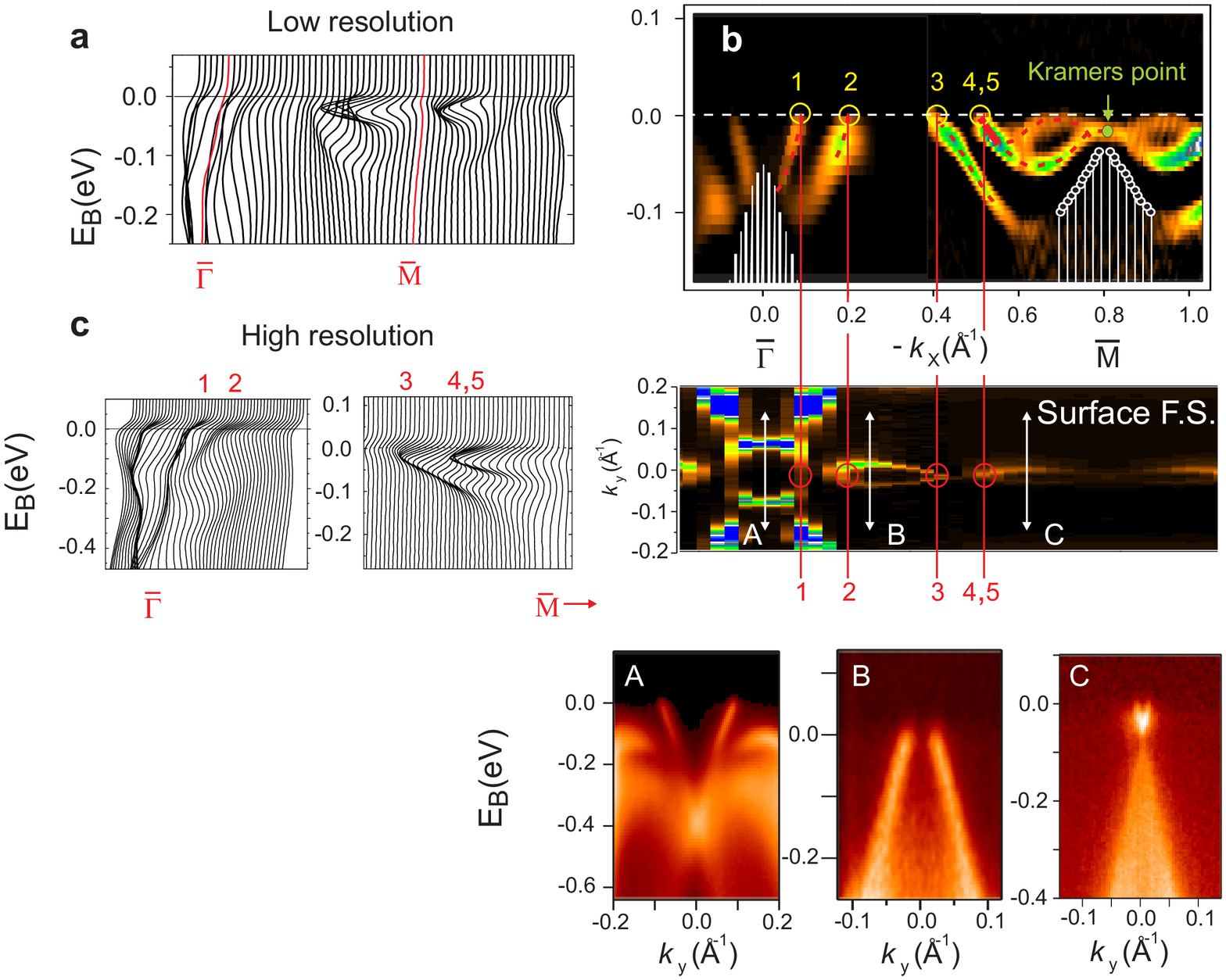}
\caption{\textbf{The Kramers' point, the gapless nature and topology
of surface states in insulating Bi$_{0.9}$Sb$_{0.1}$ is revealed
through high spatial and k-resolution ARPES.} \textbf{a,} Energy
distribution curves (EDCs) of a low resolution ARPES scan along the
surface $\bar{\Gamma}$-\={M} cut of Bi$_{0.9}$Sb$_{0.1}$.
\textbf{b,} The surface band dispersion second derivative image
along $\bar{\Gamma}$-\={M} obtained by piecing together four high
resolution ARPES scans. See main text Fig.3 for explanation of other
features. \textbf{c,} EDCs of high resolution ARPES scans in the
vicinity of surface Fermi crossings 1 and 2 and crossings 3, 4 and 5
(left panels). These crossings form the surface Fermi surface shown
in the upper right panel of \textbf{c} (see also main text Fig.2).
High resolution ARPES scans along cut directions A, B and C are
further evidence for a surface Fermi surface.}
\end{figure*}

There is a clear $k_z$ dependence of the dispersion of measured
bands A, B and C, pointing to their bulk nature. The bulk origin of
bands A, B and C is confirmed by their good agreement with tight
binding calculations (bands 3, 4 and 5 in Figs S2g and i), which
include a strong spin-orbit coupling constant of 1.5 eV derived from
bismuth$^{22}$. Band 3 drops below -5 eV at the 0.3 $XL$ point. The
slight differences between the experimentally measured band energies
and the calculated band energies at $k_y$ = 0 \AA$^{-1}$ shown in
Fig.S2f-i are due to the fact that the ARPES data were collected in
a single shot, taken in constant $\theta$ mode. This means that
electrons detected at different binding energies will have slightly
different values of $k_z$ as described in Methods, whereas the
presented tight binding calculations show all bands at a single
$k_z$. We checked that the magnitude of these band energy
differences is indeed accounted for by this explanation. Even though
the $L_a$ and $L_s$ bands in Bi$_{0.9}$Sb$_{0.1}$ are inverted
relative to those of pure semimetallic Bi, calculations show that
near $E_F$, apart from an insulating gap, they are ``mirror" bands
in terms of k dispersion (see bands 5 and 6 in Fig.S2g). Such a
close match to calculations, which also predict a linear dispersion
along the $k_z$ cut near $E_F$ (Fig.S2d), provides strong support
that the dispersion of band C, near $E_F$, is in fact linear along
$k_z$. Focusing on the $\Lambda$-shaped valence band at $L$, the
EDCs (Fig.S2a) show a single peak out to $k_y \approx \pm0.15$
\AA$^{-1}$ demonstrating that it is composed of a single band
feature. Outside this range however, an additional feature develops
on the low binding energy side of the main peak in the EDCs, which
shows up as two well separated bands in the SDI image (Fig.2f) and
signals a splitting of the band into bulk representative and surface
representative components (Fig.S2a,f). Unlike the main peak that
disperses strongly with incident photon energy, this shoulder-like
feature is present and retains the same $\Lambda$-shaped dispersion
near this k-region (open circles in Figs S2g and i) for all photon
energies used, supporting its 2D surface character. This behaviour
is quite unlike bulk band C, which attains the $\Lambda$-shaped
dispersion only near 29 eV (see main text Fig. 2b).

\section{Spin-orbit coupling is responsible for the unique Dirac-like
dispersion behaviour of the bulk bands near $E_F$}

According to theoretical models, a strongly spin-orbit coupled bulk
band structure is necessary for topological surface states to
exist$^{7-11}$. Therefore it is important to show that our
experimentally measured bulk band structure of Bi$_{0.9}$Sb$_{0.1}$
can only be accounted for by calculations that explicitly include a
large spin-orbit coupling term. As shown in the previous section,
the measured bulk band dispersion of Bi$_{0.9}$Sb$_{0.1}$ generally
follows the calculated bulk bands of pure Bi from a tight binding
model. The dispersion of the bulk valence and conduction bands of
pure bismuth near $E_F$ at the $L$ point from such a tight binding
calculation$^{22}$ with a spin-orbit coupling constant of 1.5 eV are
shown in Fig. S3b and c, which show a high degree of linearity. The
high degree of linearity can be understood from a combination of the
large Fermi velocity ($v_F$ $\approx$ 6 eV \AA\ along $k_y$) and
small inter-band (below $E_F$) gap $\Delta$ = 13.7 meV (Fig. S3).
This calculated inter-band gap of Bi (13.7 meV) is smaller than our
measured lower limit of 50 meV (main text Fig. 1a) for the
insulating gap of Bi$_{0.9}$Sb$_{0.1}$. To illustrate the importance
of spin-orbit coupling in determining the band structure near $L$,
we show the dispersion along $k_y$ and $k_z$ calculated without
spin-orbit coupling (Fig. S3d and e). While the dispersion along
$k_z$ is not drastically altered by neglecting the spin-orbit
coupling, the dispersion along $k_y$ changes from being linear to
highly parabolic. This is further evidence that our measured Dirac
point can be accounted for only by including spin-orbit coupling.
\textit{A strong spin-orbit coupling constant acts as an internal
quantizing magnetic field for the electron system$^6$ which can give
rise to a quantum spin Hall effect without any externally applied
magnetic field$^{3,4,5,12,37}$. Therefore, the existence or the
spontaneous emergence of the surface or boundary states does not
require an external magnetic field.}

\section{Matching the surface state Fermi crossings and the topology
of the surface Fermi surface in bulk insulating
B\lowercase{i}$_{0.9}$S\lowercase{b}$_{0.1}$}

In order to count the number of singly degenerate surface state
Fermi crossings$^{24,28,38}$ along the $\bar{\Gamma}$-\={M} cut of
the surface BZ, high photon energy ARPES scans, which allow mapping
of the entire k range from $\bar{\Gamma}$-\={M} to fall within the
detector window at the expense of lower instrument resolution, were
taken to preliminarily identify the k-space locations of the Fermi
crossings (Fig. S4a). Having determined where these surface state
Fermi crossings lie in k-space, we performed several high resolution
ARPES scans, each covering a successive small k interval in the
detector window, in order to construct a high resolution band
mapping of the surface states from $\bar{\Gamma}$ to \={M}. The
second derivative image of the surface band dispersion shown in
Fig.S4b was constructed by piecing together four such high
resolution scans. Fig.S4c shows energy distribution curves of high
resolution ARPES scans in the vicinity of each surface Fermi
crossing, which together give rise to the surface Fermi surface
shown. No previous work$^{24,26-30,35,36}$ has reported the band
dispersion near the $L$-point (thus missing the Dirac bands) or
resolved the Kramers point near the \={M} point, which is crucial to
determine the topology of the surface states. For this reason there
is no basis for one-to-one comparison with previous work, since no
previous ARPES data exists in the analogous k-range. Note that
surface band dispersions along the cuts A, B and C are highly
linear. This is indirect evidence for the existence of the bulk
Dirac point since surface states are formed when the bulk state wave
functions are subjected to the boundary conditions at the cleaved
plane.


\begin{thebibliography}{h}

\bibitem{Klitzing}
von Klitzing, K., Dorda, G. \& Pepper, M. New method for
high-accuracy determination of the fine-structure constant based on
quantized Hall resistance. \textit{Phys. Rev. Lett}. \textbf{45},
494-497 (1980).

\bibitem{Tsui}
Tsui, D.C., Stormer, H.L. \& Gossard, A.C. Two-dimensional
magnetotransport in the extreme quantum limit. \textit{Phys. Rev.
Lett}. \textbf{48}, 1559-1562 (1982).

\bibitem{Kane(Graphene)}
Kane, C.L. \& Mele, E.J. Quantum spin Hall effect in graphene.
\textit{Phys. Rev. Lett}. \textbf{95}, 226801 (2005).

\bibitem{Bernevig(QSHE)}
Bernevig, B.A. \& Zhang, S.-C. Quantum spin Hall effect.
\textit{Phys. Rev. Lett.} \textbf{96}, 106802 (2006).

\bibitem{Sheng(QSHE)}
Sheng, D.N., Weng, Z.Y., Sheng, L. \& Haldane, F.D.M. Quantum spin
Hall effect and topological Chern numbers. \textit{Phys. Rev. Lett.}
\textbf{97}, 036808 (2006).

\bibitem{Haldane(P-anomaly)}
Haldane, F.D.M. Model for a quantum Hall effect without Landau
levels: Condensed-matter realization of the ``parity anomaly".
\textit{Phys. Rev. Lett.} \textbf{61}, 2015-2018 (1988).

\bibitem{Fu:STI1}
Fu, L. \& Kane, C.L. Topological insulators with inversion symmetry.
\textit{Phys. Rev.} \textbf{B76}, 045302 (2007).

\bibitem{Murukami}
Murakami, S. Phase transition between the quantum spin Hall and
insulator phases in 3D: emergence of a topological gapless phase.
\textit{New. J. Phys.} \textbf{9}, 356 (2007).

\bibitem{Fu:STI2}
Fu, L., Kane, C.L. \& Mele, E.J. Topological insulators in three
dimensions. \textit{Phys. Rev. Lett}. \textbf{98}, 106803 (2007).

\bibitem{Moore:STI1}
Moore, J.E. \& Balents, L. Topological invariants of
time-reversal-invariant band structures. \textit{Phys. Rev.}
\textbf{B75}, 121306(R) (2007).

\bibitem{Roy}
Roy, R. Three dimensional topological invariants for time reversal
invariant Hamiltonians and the three dimensional quantum spin Hall
effect. Preprint at
$\langle$http://arxiv.org/abs/cond-mat/0607531$\rangle$ (2006).

\bibitem{Bernevig:HgTe}
Bernevig, B.A., Hughes, T.L. \& Zhang, S.-C. Quantum spin Hall
effect and topological phase transition in HgTe quantum wells.
\textit{Science} \textbf{314}, 1757-1761 (2006).

\bibitem{Konig}
K\"{o}nig, M. \textit{et al}. Quantum spin Hall insulator state in
HgTe quantum wells. \textit{Science} \textbf{318}, 766-770 (2007).

\bibitem{Wolff}
Wolff, P.A. Matrix elements and selection rules for the two-band
model of bismuth. \textit{J. Phys. Chem. Solids} \textbf{25},
1057-1068 (1964).

\bibitem{Fukuyama}
Fukuyama, H. \& Kubo, R. Interband effects of magnetic
susceptibility. II. Diamagnetism of bismuth. \textit{J. Phys. Soc.
Jpn.} \textbf{28}, 570-581 (1970).

\bibitem{Buot}
Buot, F.A. Weyl transformation and the magnetic susceptibility of a
relativistic Dirac electron gas. \textit{Phys. Rev.} \textbf{A8},
1570-1581 (1973).

\bibitem{Lenoir}
Lenoir, B. \textit{et al}. Bi-Sb alloys: an update.
\textit{Fifteenth International Conference on Thermoelectrics}, 1-13
(1996).

\bibitem{Zhang}
Zhang, Y. \textit{et.al}. Experimental observation of the quantum
Hall effect and Berry's phase in graphene. \textit{Nature}
\textbf{438}, 201-204 (2005).

\bibitem{Novoselov}
Novoselov, K.S. \textit{et al.} Room temperature quantum Hall effect
in graphene. \textit{Science} \textbf{315}, 1379 (2007).

\bibitem{Zhou}
Zhou, S.Y. \textit{et al}. Substrate-induced bandgap opening in
epitaxial graphene. \textit{Nature Mat.} \textbf{6}, 770-775 (2007).

\bibitem{Behnia}
Behnia, K., Balicas, L. \& Kopelevich, Y. Signatures of electron
fractionalization in ultraquantum bismuth. \textit{Science}
\textbf{317}, 1729-1731 (2007).

\bibitem{Liu}
Liu, Y. \& Allen, E. Electronic structure of semimetals Bi and Sb.
\textit{Phys. Rev.} \textbf{B52}, 1566-1577 (1995).

\bibitem{Hebel}
Hebel, L.C. \& Smith, G.E. Interband transitions and band structure
of a BiSb alloy. \textit{Phys. Lett.} \textbf{10}, 273-275 (1964).

\bibitem{Kane(QSHE-Z2)}
Kane, C.L. \& Mele, E.J. $Z_2$ topological order and the quantum
spin Hall effect. \textit{Phys. Rev. Lett}. \textbf{95}, 246802
(2005).

\bibitem{Ast:Bi1}
Ast, C.R. \& Hochst, H. Fermi Surface of Bi(111) Measured by
Photoemission Spectroscopy. \textit{Phys. Rev. Lett.} \textbf{87},
177602 (2001).

\bibitem{Hochst}
Hochst, H. \& Gorovikov, S. Lack of electron-phonon coupling along
two-dimensional bands in Bi$_{1-x}$Sb$_x$ single crystal alloys.
\textit{J. Elect. Spectrosc. Relat. Phenom.} \textbf{351}, 144-147
(2005). This work does not measure the surface state along the
critical $\bar{\Gamma}-\bar{M}$ direction or detect the bulk Dirac
spectrum near L.

\bibitem{Hofmann}
Hofmann, P. The surfaces of bismuth: Structural and electronic
properties. \textit{Prog. Surf. Sci.} \textbf{81}, 191-245 (2006).

\bibitem{Hirahara}
Hirahara, T. \textit{et al}. Direct observation of spin splitting in
bismuth surface states. \textit{Phys. Rev.} \textbf{B76}, 153305
(2007).

\bibitem{Hengsberger}
Hengsberger, M. \textit{et al}. Photoemission study of the carrier
bands in Bi(111). \textit{Eur. Phys. J.} \textbf{17}, 603-608
(2000).

\bibitem{Ast:Bi2}
Ast, C.R. \& Hochst, H. Electronic structure of a bismuth bilayer.
\textit{Phys. Rev.} \textbf{B67}, 113102 (2003).

\bibitem{31}
Kopelevich, Y. $et$ $al.$ Universal magnetic-field-driven
metal-insulator-metal transformations in graphite and bismuth.
\textit{Phys. Rev.} \textbf{B 73}, 165128 (2006).

\bibitem{32}
Hufner, S. Photoelectron Spectroscopy (Springer, Berlin, 1995).

\bibitem{33}
Novoselov, K. S. $et$ $al$. Two-dimensional gas of massless Dirac
fermions in graphene. \textit{Nature} \textbf{438}, 197-200 (2005).

\bibitem{34}
Bostwick, A., Ohta, T., Seyller, T., Horn, K. \& Rotenberg, E.
Quasiparticle dynamics in graphene. \textit{Nature Phys.}
\textbf{3}, 36-40 (2007).

\bibitem{35}
Jezequel, G., Thomas, J. \& Pollini, I. Experimental band structure
of semimetal bismuth. \textit{Phys. Rev.} \textbf{B 56}, 6620-6626
(1997).

\bibitem{36}
Ast, C. R. \& Hochst, H. High-resolution mapping of the
three-dimensional band structure of Bi(111). \textit{Phys. Rev.}
\textbf{B 70}, 245122 (2004).

\bibitem{37}
Sheng, L., Sheng, D. N., Ting, C. S. \& Haldane, F. D. M.
Nondissipative spin Hall effect via quantized edge transport.
\textit{Phys. Rev. Lett}. \textbf{95}, 136602 (2005).

\bibitem{38}
Kim, T. K. $et$ $al$. Evidence against a charge density
wave on Bi(111). \textit{Phys. Rev}. \textbf{B 72}, 085440 (2005).

\bibitem{Science}
Hsieh, D., Hasan, M.Z. \textit{et al}. Observation of unconventional
quantum spin textures in topological insulators. \textit{Science}
\textbf{323}, 919 (2009).

\end{thebibliography}

\end{document}